\documentclass[11pt, draftclsnofoot, onecolumn]{IEEEtran}

\usepackage{cite,graphicx,amsmath,amssymb}
\usepackage{subfigure}
\usepackage{citesort}
\usepackage{fancyhdr}
\usepackage{mdwmath}
\usepackage{mdwtab}
\usepackage{balance}
\usepackage{xcolor}
\usepackage{bm}
\usepackage{amsthm}
\usepackage{threeparttable}
\usepackage{algorithm}
\usepackage{algorithmic}
\usepackage{multirow}
\usepackage{flafter}
\usepackage{makecell}
\usepackage{diagbox}

\providecommand{\url}[1]{#1}
\begin{document}
\title{Reconfigurable Intelligent Surface (RIS) Aided Multi-User Networks: Interplay Between NOMA and RIS}
% \markboth{\textit{A Manuscript Submitted to The IEEE   Communications Magazine} }

\author{
 Yuanwei~Liu, Xidong Mu, Xiao Liu, Marco Di Renzo, Zhiguo Ding, and Robert Schober
%~\IEEEmembership{Member,~IEEE,}
%\IEEEauthorblockN{,~\IEEEmembership{Member,~IEEE,}}
% ,~\IEEEmembership{Fellow,~IEEE,}
%and \IEEEauthorblockN{,~\IEEEmembership{Senior Member,~IEEE,}
% }
%\thanks{Y. Liu and X. Liu are with Queen Mary University of London; X. Mu is with Beijing University of Posts and Telecommunications; M. Di. Renzo is with Universit\'e Paris-Saclay; Z. Ding is with the The University of Manchester; R. Schober is with Friedrich-Alexander-University Erlangen-N{\"u}rnberg (FAU).}
\thanks{Y. Liu and X. Liu are with Queen Mary University of London, London E1 4NS, UK, (email: \{yuanwei.liu, x.liu, x.gao\}@qmul.ac.uk).}
\thanks{X. Mu is with Beijing University of Posts and Telecommunications, Beijing 100876, China (email: muxidong@bupt.edu.cn).}
\thanks{M. Di Renzo is with Universit\'e Paris-Saclay, CNRS, CentraleSup\'elec, Laboratoire des Signaux et Syst\`emes, 3 Rue Joliot-Curie, 91192 Gif-sur-Yvette, France. (email: marco.direnzo@centralesupelec.fr).}
\thanks{Z. Ding is with the School of Electrical and Electronic Engineering, The University of Manchester, Manchester, UK (e-mail:
zhiguo.ding@manchester.ac.uk).}
\thanks{R. Schober is with the Institute for Digital Communications, Friedrich-Alexander-University Erlangen-N{\"u}rnberg (FAU), Germany (e-mail: robert.schober@fau.de).}
}

\maketitle

\begin{abstract}
This article focuses on the exploitation of reconfigurable intelligent surfaces (RISs) in multi-user networks employing orthogonal multiple access (OMA) or non-orthogonal multiple access (NOMA), with an emphasis on investigating the interplay between NOMA and RIS. Depending on whether the RIS reflection coefficients can be adjusted only once or multiple times during one transmission, we distinguish between \emph{static} and \emph{dynamic} RIS configurations. In particular, the capacity region of RIS-aided single-antenna NOMA networks is discussed and compared with the OMA rate region from an information-theoretic perspective, revealing that the dynamic RIS configuration is capacity-achieving. Then, the impact of the RIS deployment location on the performance of different multiple access schemes is investigated, which reveals that asymmetric and symmetric deployment strategies are preferable for NOMA and OMA, respectively. Furthermore, for RIS-aided multiple-antenna NOMA networks, three novel joint active and passive beamformer designs are proposed based on both beamformer-based and cluster-based strategies. Finally, open research problems for RIS-NOMA networks are highlighted.
\end{abstract}

\section{Introduction}
Reconfigurable intelligent surfaces (RISs)~\cite{Renzo2020JSAC,Huang_Holographic,IRS_WU} have been proposed as a promising candidate technology for next-generation wireless networks due to their potential to control the propagation of wireless signals through customized reflections and refractions~\cite{Renzo2020JSAC}. More particularly, by adjusting the phases and amplitudes of the reconfigurable RIS elements, the desired signals can be enhanced and the undesired interfering signals can be mitigated to realize a so-called `Smart Radio Environment'~\cite{Renzo2020JSAC,di2019smart}. As a result, the coverage and spectral efficiency of existing wireless networks can be significantly improved. In contrast to conventional active relaying technologies, RISs consume much less energy and are less expensive since they do not employ radio frequency (RF) chains or high power consuming components. RISs can be seamlessly integrated into existing wireless communication networks to realize different objectives, including the improvement of the energy efficiency~\cite{Huang_EE} and the mitigation of inter-user interference~\cite{Wu_MAG}. Motivated by these benefits of RISs, we focus our attention on investigating RIS-aided multi-user networks. Compared to RIS-aided single-user networks, simultaneously serving multiple users in the time/frequency/power domains introduces new challenges for smartly coordinating the interference so as to efficiently exploit the spatial degrees of freedom (DoFs) facilitated by the active and passive beamforming at the base station (BS) and the RIS, respectively. To begin with, we provide a brief introduction to the integration of RISs into multi-user networks and the combination of multiple access (MA) schemes with two types of RIS configurations.
\subsection{RIS Aided Multi-User Networks Exploiting Different MA Schemes}
\begin{figure*}
  \centering
  \includegraphics[width=6in]{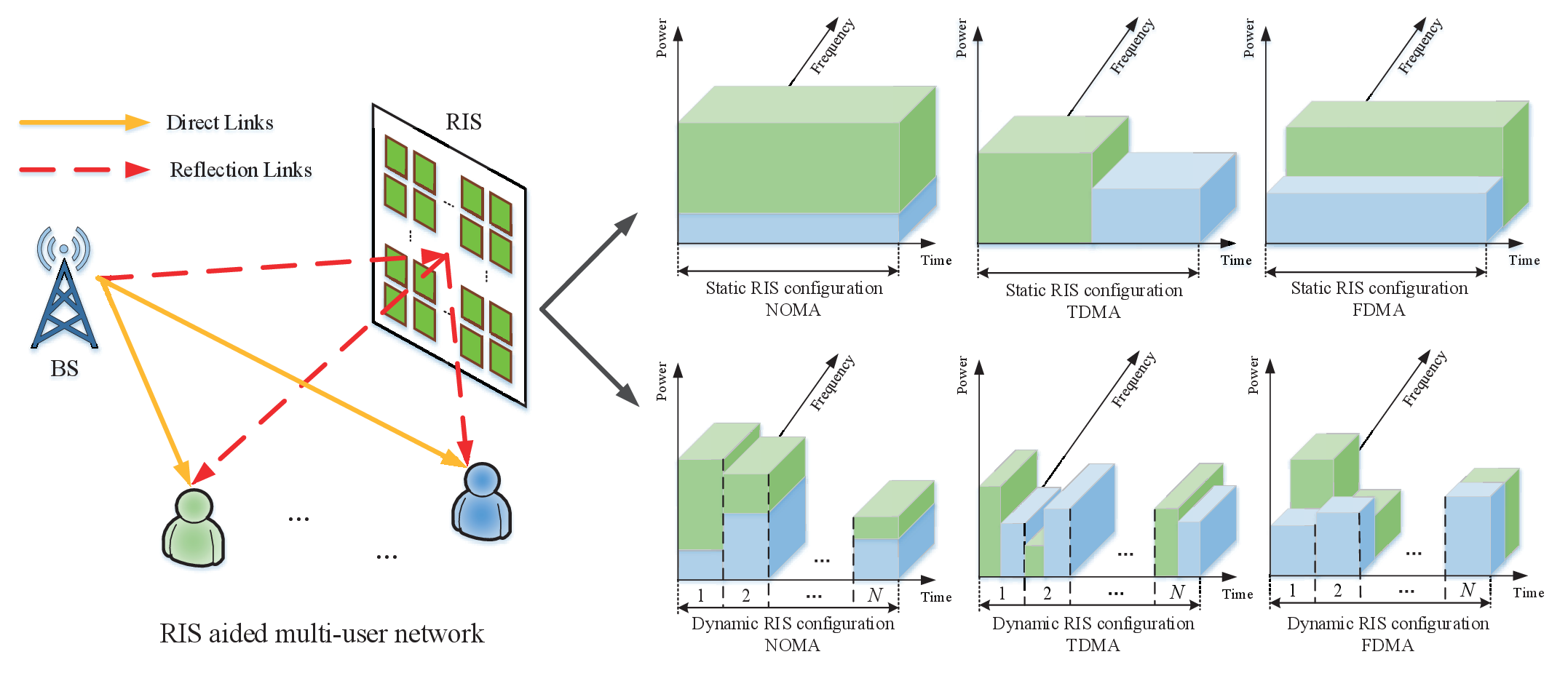}\\
  \caption{Illustration of an RIS-aided multi-user network employing NOMA, TDMA, and FDMA for static and dynamic RIS configurations.}\label{configurations}
\end{figure*}%19
Fig. \ref{configurations} illustrates an RIS-aided multi-user network employing non-orthogonal multiple access (NOMA) and orthogonal multiple access (OMA), respectively. Specifically, time division multiple access (TDMA) and frequency division multiple access (FDMA) is analyzed for OMA. As for the RIS configuration, two options are considered, namely a \emph{static RIS configuration} and a \emph{dynamic RIS configuration}. As for the static RIS configuration, the reflection coefficients of the RIS can be adjusted only once during one transmission. Then, the users are distinguished in the power/time/frequency domain based on a static MA strategy, as illustrated at the top right of Fig. \ref{configurations}. As for the dynamic RIS configuration, the RIS reflection coefficients are reconfigured $N$ times during one transmission and the time durations of all configurations are assumed to have equal size. As a result, the RIS can create an artificial fading channel. This unique feature enabled by the RIS allows dynamic resource allocation for the adopted MA scheme, as illustrated at the bottom right of Fig. \ref{configurations}. The dynamic RIS configuration is a viable option especially for static or quasi-static communication scenarios (i.e., the users are static or moving slowly), which are the most typical application scenarios of RISs due to their limited coverage. According to \cite{RFfocus}, the time consumption for reconfiguring an RIS is 0.22 millisecond (ms), which is much shorter than the typical channel coherence time (e.g., tens of ms), thus facilitating the realization of the dynamic RIS configuration.

\subsection{Motivation for Applying NOMA in RIS-aided Multi-user Networks}

Inspired by the aforementioned considerations for MA strategies in RIS-aided wireless networks, we focus our attention on NOMA. On the one hand, NOMA constitutes an efficient MA strategy in RIS-aided multi-user networks, thus improving the spectral efficiency and enhancing the connectivity. On the other hand, RISs introduce the following benefits in existing NOMA networks: 1) The reflection links of RISs can enhance the performance of existing NOMA networks by providing additional signal diversity without requiring extra time slots and energy. 2) RISs increase the design flexibility of NOMA networks, leading to the transition from `\emph{channel condition-based NOMA}' to `\emph{quality of service (QoS)-based NOMA}'. More specifically, in conventional NOMA networks, the successive interference cancelation (SIC) decoding order of the users is generally determined by the given `dumb' channel conditions, which are determined by the environment. As a `channel changing' technique, RISs can enhance or degrade the channel quality of individual users by adjusting the reflection coefficients and the deployment locations of the RISs. This enables a `\emph{smart}' NOMA design. 3) The use of RISs can reduce the constraints in multiple-antenna NOMA networks. In conventional multiple-input multiple-output (MIMO) NOMA networks, some constraints on the numbers of antennas at the transmitters and receivers may need to be satisfied~\cite{Liu2018Multiple}. With the aid of RISs, such constraints can be relaxed due to the additional passive array gains provided by the RISs. Given the potential benefits introduced by the interplay between NOMA and RISs, researchers have begun to investigate RIS-NOMA networks. Nevertheless, the application of NOMA in RIS-aided networks give rise to numerous new challenges, which provide the main motivation for this article.

Our main contributions can be summarized as follows.
\begin{itemize}
  \item The fundamental information-theoretic capacity/rate limits of RIS-aided multi-user single-antenna networks are discussed and compared for NOMA and OMA employing different RIS configurations. Notably, the dynamic RIS configuration is shown to be capacity-achieving.
  \item The deployment design of RISs is discussed and preferable deployment strategies for different MA schemes are presented.
  \item Novel \textcolor{black}{joint active and passive beamformer designs} are proposed for RIS-aided multiple-antenna NOMA networks. Specifically, beamformer-based strategies and cluster-based strategies are developed. %For each design, the advantages and limitations are also discussed.
\end{itemize}
\section{Capacity Limits: An Information-Theoretic Perspective}
In this section, we discuss the information-theoretic capacity limits of the RIS-aided multi-user broadcast channel (BC). Notably, we reveal that NOMA with a dynamic RIS configuration maximizes the capacity region and is a capacity-achieving transmission strategy.

We focus on the RIS-aided BC, where one single-antenna base station (BS) serves $K$ single-antenna users assisted by an RIS, as illustrated in Fig. \ref{configurations}. It is known that the capacity region of the BC with a single-antenna transmitter and single-antenna receivers is achieved by the NOMA transmission scheme, i.e., by invoking superposition coding (SC) at the transmitter and SIC at the receivers. The capacity region contains all the achievable rate tuples of the users under a total transmit power constraint. In contrast to the conventional BC without an RIS, the reflection coefficients at the RIS and the wireless resource allocation at the BS have to be jointly optimized for obtaining the capacity region of the considered RIS-aided BC. Next, we discuss the capacity region of the RIS-aided BC for static and dynamic RIS configurations, respectively.
\subsection{Static RIS Configuration}
In this case, the reflection coefficients of the RIS are fixed for the entire transmission, as illustrated in Fig. \ref{configurations}. By employing fixed the reflection coefficients, the considered RIS-aided BC is identical to the conventional BC. As a result, the capacity region of the RIS-aided BC based on the static RIS configuration can be obtained by exhaustively considering all possible reflection coefficients, and taking the union of all achievable NOMA rate tuples~\cite{Mu_Capacity}. The size of the search space is determined by the number of reflection elements and the resolution at which the phases of the RIS elements can be adjusted. Similarly, the rate regions achieved by the TDMA and FDMA schemes can be obtained based on the achievable rate tuples for each scheme.
\subsection{Dynamic RIS Configuration}
As for the dynamic RIS configuration, the RIS reflection coefficients are assumed to be reconfigured $N$ times during one transmission instance, as illustrated in Fig. \ref{configurations}. The capacity region that corresponds to the dynamic RIS configuration is obtained by the union of all achievable NOMA rate tuples over all the possible combinations of $N$ values of reflection coefficients for each RIS element. However, determining this capacity region is prohibitively complex since the size of the set of all possible combinations of reflection coefficients grows exponentially with $N$.

To circumvent this issue and to determine a performance upper bound for the capacity region, we consider the ideal case when $N$ tends to infinity, which implies that the finite time duration that is needed for configuring and updating the configuration of the RIS is neglected, i.e., the RIS configuration can be updated in real time. In this case, time sharing can be carried out among all possible reflection coefficients. Thus, the capacity region for $N$ tending to infinity can be obtained by taking the convex hull of the capacity regions achieved by all possible static RIS configurations~\cite{Mu_Capacity}. Moreover, it can be verified that the corresponding capacity region contains the capacity region for finite values of $N$, which in turn contains the capacity region of the static RIS configuration. Thus, the dynamic RIS configuration with NOMA is capacity-achieving for the considered RIS-aided multi-user BC~\cite{Mu_Capacity}. The TDMA/FDMA rate region that corresponds to the dynamic RIS configuration can be obtained in a similar manner.
\subsection{Discussions and Insights}
\begin{figure}[!t]
  \centering
  \includegraphics[width=3in]{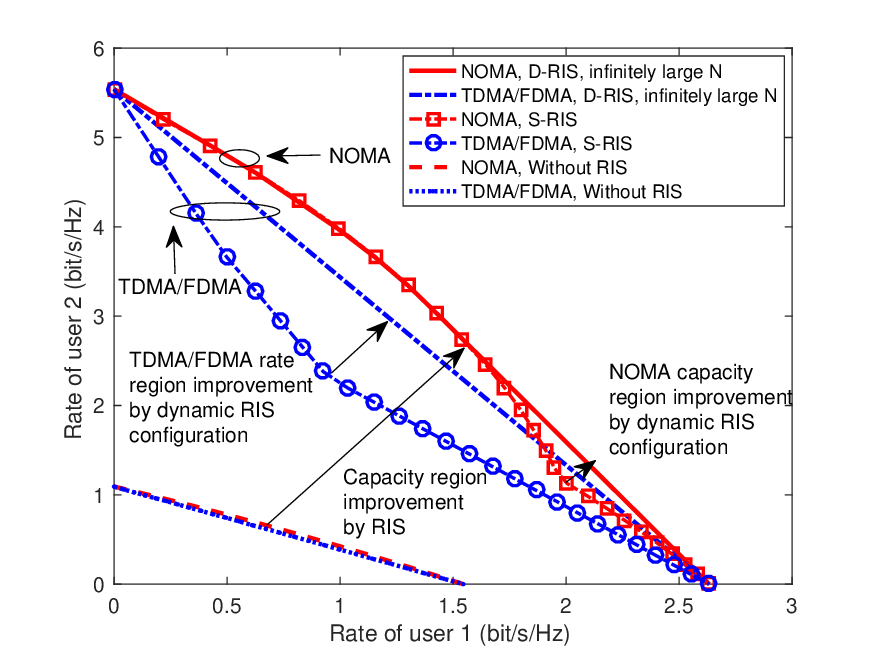}\\
  \caption{\textcolor{black}{Illustration of the boundaries of the NOMA capacity region and the TDMA/FDMA rate region for an RIS-aided two-user BC employing static and dynamic RIS configurations and for the system without RIS. The system parameter settings and path loss models can be found in~\cite[Section V]{Mu_Capacity}. Here, ``D-RIS'' and ``S-RIS'' refer to the dynamic RIS configuration and the static RIS configuration, respectively.}}\label{capacity}
\end{figure}%66
To demonstrate the fundamental capacity/rate limits mentioned above, we consider a numerical example for an RIS-aided two-user BC. We assume that the direct links and the reflection links are subject to Rayleigh and Rician fading, respectively. For a random channel realization, we compare in Fig. \ref{capacity} the NOMA capacity region of the RIS-aided two-user BC with the rate region achieved by the TDMA/FDMA schemes under the considered static and dynamic RIS configurations~\cite{Mu_Capacity}. As for the dynamic RIS configuration, we evaluate the case when $N$ tends to infinity. \textcolor{black}{In this case, the theoretical analysis conducted in \cite{Mu_Capacity} revealed that the capacity-achieving transmission strategy for NOMA is to transmit in an alternating manner to different user groups using different decoding orders.} As expected, the NOMA capacity region contains the TDMA/FDMA rate region for both RIS configurations. Moreover, we observe that the capacity/rate gain introduced by dynamically adjusting the phases of the RIS elements is much more pronounced for TDMA/FDMA than for NOMA. This highlights an interesting capacity-complexity tradeoff between TDMA/FDMA and NOMA: Although NOMA introduces additional complexity for applying SIC at the users, the implementation complexity of the RIS is reduced, since the dynamic configuration is generally not needed as it results in only a small gain compared with the static RIS configuration. To fully reap the RIS benefits, on the other hand, OMA schemes require a much more sophisticated RIS design than NOMA, i.e., real-time RIS control is needed in order to realize the dynamic RIS configuration. On the other hand, SIC is avoided in OMA due to the orthogonal transmission. The tradeoff between implementation complexity and network capacity is an interesting topic for further investigation. \textcolor{black}{To assess the benefits of deploying RISs, Fig. \ref{capacity} also contains the NOMA capacity region and the TDMA/FDMA rate region in the absence of an RIS. As can be seen from Fig. \ref{capacity}, for both RIS configurations, a significant capacity/rate gain is introduced by deploying an RIS. Furthermore, the capacity gain of NOMA over OMA in the RIS-aided two-user BC is much more pronounced compared to the corresponding gain if an RIS is not deployed. This underscores the benefits of employing RISs in NOMA networks.}

In this section, the capacity limits for the RIS-aided single-antenna BC were discussed. The capacity limits and the capacity-achieving strategies for RIS-aided networks with multiple-antenna transmitters and receivers are still largely unknown and constitute an interesting topic for future research.

\section{RIS Deployment Design for Multiple Access}

Having discussed the fundamental capacity limits of RIS-aided multi-user networks, we focus our attention on the optimal deployment of RIS for different MA schemes. Besides the optimization of the reflection coefficients, the optimization of the locations of the RIS is another DoF that, when applicable, has an impact on the system performance. This is because the path loss of the RIS-assisted link depends on the product of the distances of the BS-RIS link and the RIS-user link~\cite{Ozdogan_pathloss}. Therefore, it is necessary to jointly consider the optimization of the location of the RIS and the optimal allocation of the wireless resources to users in order to take advantage of the presence of RISs. In this section, we provide some initial results on the joint optimization of the RIS deployment and the MA scheme for RIS-aided multi-user networks.

We first discuss the interplay between RIS deployment and NOMA for enabling a `QoS-based NOMA' operation. For ease of exposition, we consider a two-user NOMA network. In conventional NOMA networks without RISs, the user with the strongest direct channel gain is designated as the strong user, and the other user as the weak user. In RIS-aided NOMA networks, two deployment scenarios are of particular interests.
\begin{itemize}
  \item \textbf{Consolidation}: The RIS can be deployed in the vicinity of the strong user to further enhance its channel quality. By doing so, the channel disparity between the two users can be further enlarged, and NOMA yields higher performance gains over OMA. This consolidation-based RIS deployment strategy is suitable when the strong user has a stricter QoS requirement than the weak user.
  \item \textbf{Reverse}: The RIS can be deployed near the weak user to enhance its channel quality and to reverse the original `channel-condition based' decoding order in the absence of the RIS. The reverse-based RIS deployment strategy is appealing when the weak user has a higher QoS requirement than the strong user in the system without RIS.
\end{itemize}

As for TDMA and FDMA networks, the RIS can be deployed in the vicinity of the users experiencing poor channel conditions in order to enhance their performance. Another effective RIS deployment strategy for TDMA/FDMA networks is based on the users' spatial distribution. For instance, deploying the RIS in the vicinity of a cluster of users with similar channel conditions can jointly improve their communication performance.

\begin{figure}[!htb]
  \centering
  \includegraphics[width=3in]{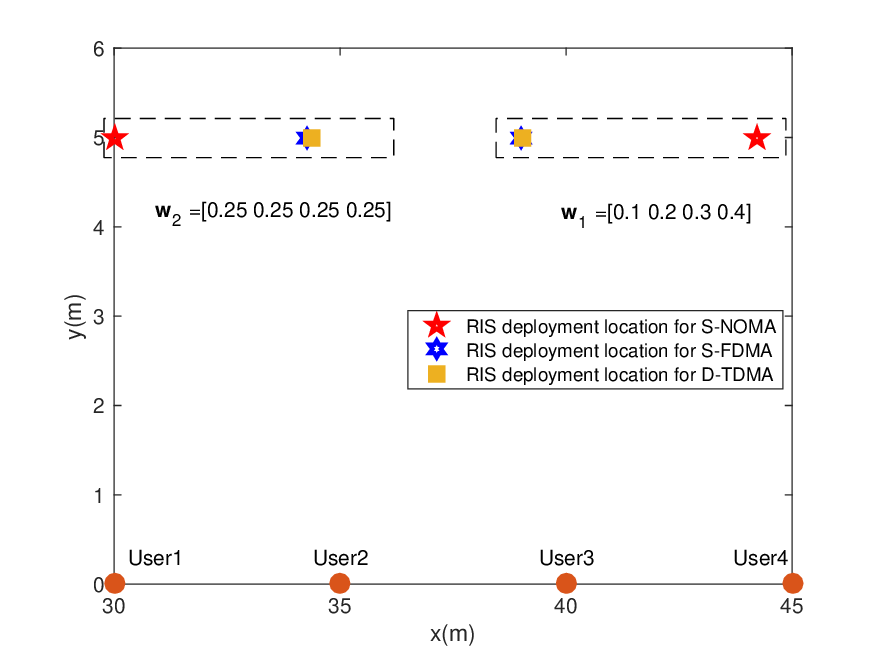}\\
  \caption{Illustration of the optimal RIS deployment for different MA schemes based on the optimization of the weighted sum-rate for two different user rate weight vectors ${\mathbf{w}_1}$ and ${\mathbf{w}_2}$. The BS is located at $\left( {0,0,5} \right)$ m. The heights of the RIS and the users are $1.5$ m and $0$ m, respectively. The location of the RIS is varied between $30$ m and $45$ m along the $x$-axis. The setup of the other parameter settings can be found in~\cite[Section VI]{Mu_Deployment}.}\label{Deployment_MA}
\end{figure}%75
Fig. \ref{Deployment_MA} illustrates for different MA schemes the optimal locations of the RIS that maximize the weighted sum-rate of a four-user network~\cite{Mu_Deployment}. The simulation results are obtained by assuming that the BS-user links are blocked and only the BS-RIS-user links are available and are subject to Rician fading. The NOMA and FDMA schemes with a static RIS configuration, denoted by S-NOMA and S-FDMA, respectively, and the TDMA scheme with a dynamic RIS configuration, denoted by D-TDMA, are illustrated. As far as the D-TDMA scheme is concerned, each user is served sequentially by setting different RIS reflection coefficients exploiting the dynamic RIS configuration. Fig. \ref{Deployment_MA} illustrates that an \emph{asymmetric} RIS deployment strategy (i.e., deploying the RIS close to one of the users and far from the other users) is preferable for NOMA, while a \emph{symmetric} RIS deployment strategy (i.e., deploying the RIS at similar distances from all the users) provides a better weighted sum-rate in both FDMA and TDMA networks regardless of the type of RIS configuration. This behaviour is obtained because an asymmetric RIS deployment strategy leads to pronounced channel differences between the users, which is beneficial for the operation of NOMA. \textcolor{black}{To elaborate, let us consider the scenario with the non-uniform user rate weight vector. In this case, the RIS is deployed closest to user 4 who has the highest rate weight. This deployment, therefore, enhances the channel quality of user 4, while degrading the channel quality of the other users. As a result, the distinctiveness of the channel of one user is increased, which is beneficial for the application of NOMA.} By contrast, similar channel conditions are preferable for FDMA and TDMA, which leads to a symmetric RIS deployment.

For the joint design of RIS deployment and MA scheme, many open research problems need to be tackled. For example, since the users generally move in the network, dynamically optimizing the location of the RIS (e.g., by mounting RISs on unmanned vehicles) and selecting the most appropriate MA and RIS configurations constitutes a promising approach for satisfying the time-variant QoS requirements of the users. However, conventional standard optimization methods may be ineffective for handling the resulting complicated optimization problems. In this context, the application of machine learning (ML) based on a long-term performance metric can be a suitable solution.

\section{RIS-aided Multiple-Antenna NOMA Networks: \textcolor{black}{Joint Active and Passive Beamformer Design}}
In this section, we consider an RIS-aided multiple-antenna NOMA network, in which the transmitters and the receivers are equipped with multiple antennas to further enhance the spectral efficiency. Depending on whether one beamformer is used to serve one user or multiple users, we broadly classify current multiple-antenna NOMA transmission schemes into two categories, namely beamformer-based strategies and cluster-based strategies~\cite{Liu2018Multiple}. For each strategy, we discuss \textcolor{black}{joint active and passive beamformer designs} for RIS-aided multiple-antenna NOMA networks.

\subsection{Beamformer-based RIS-NOMA: A Novel Equivalent Reconfigurable Channel Inspired Design}
\begin{figure*}[!htb]
\centering
\subfigure[]{\label{BB structure}
\includegraphics[width= 4in]{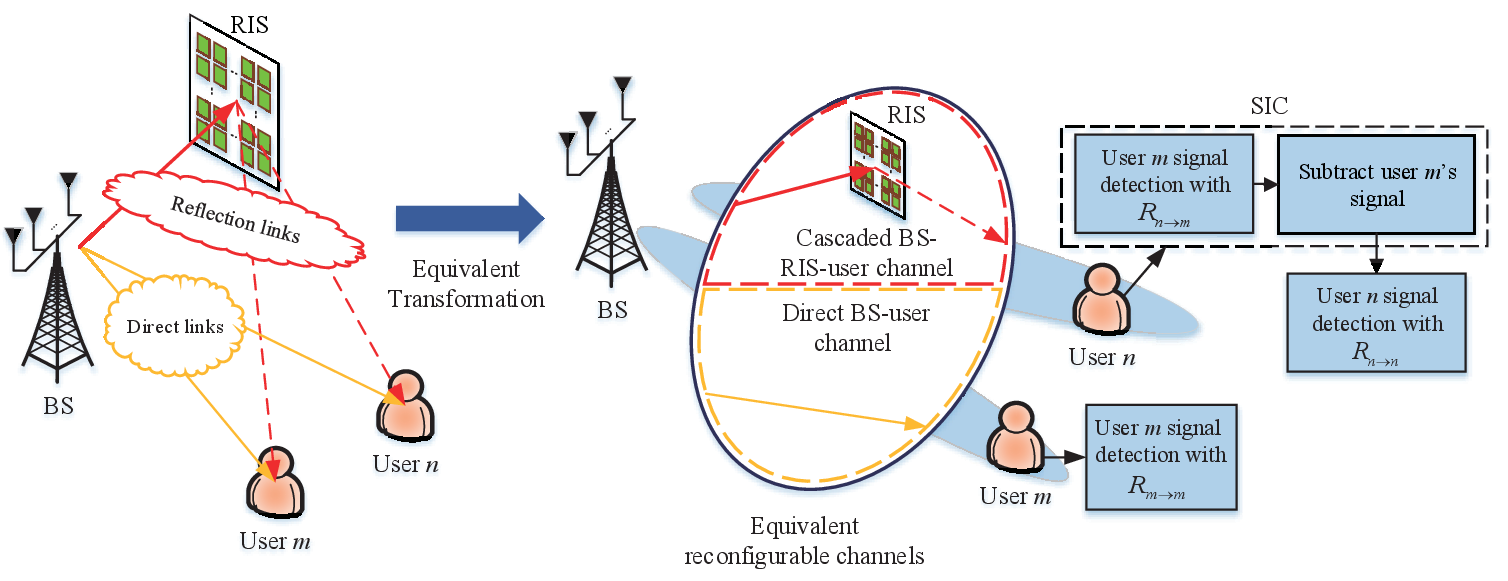}}
\subfigure[]{\label{NDQ}
\includegraphics[width= 2in]{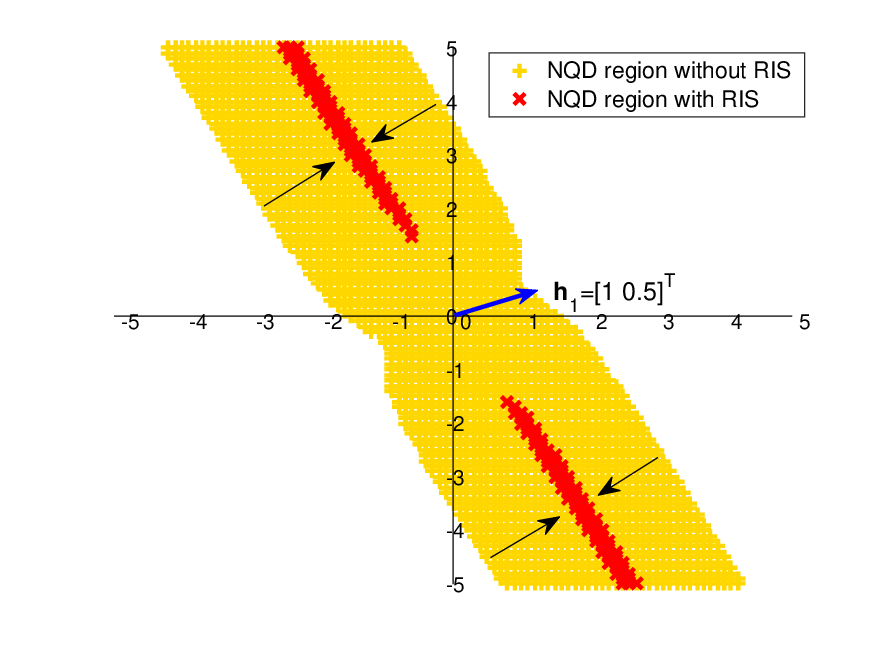}}
\caption{\textcolor{black}{Beamformer-based strategy for RIS-aided multiple-antenna NOMA networks: a novel equivalent reconfigurable channel inspired design and initial numerical results. (a) Equivalent reconfigurable channel transformation. (b) Reduction of NDQ region by deploying the RIS, where a two-antenna BS serves two single-antenna users with the aid of a 20-element RIS. All the users have a rate requirement of 1 bit/s/Hz. The channel vector of user 1 is fixed at ${{\mathbf{h}}_1}$ and the BS-RIS-user links are randomly generated. Here, ``NQD'' refers to non-quasi-degradation.}}\label{Beamformer-based}
\end{figure*}

%\begin{figure*}[!htb]
%  \centering
%  \includegraphics[width=5in]{beamformer_based_RIS_combine2.eps}\\
%  \caption{Beamformer-based strategy for RIS-aided multiple-antenna NOMA networks: a novel equivalent reconfigurable channel inspired design.}\label{Beamformer-based}
%\end{figure*}
The introduction of RISs poses new challenges for the design of beamformer-based multiple-antenna NOMA strategies, as the channel quality of the users is determined by both the direct links and the RIS-assisted links, as shown on the left hand side of Fig. \ref{BB structure}. \textcolor{black}{In order to address this issue, we propose a novel equivalent reconfigurable channel inspired design, where the equivalent channel gains facilitated by combining the direct links and the RIS-assisted links are exploited for jointly optimizing the active beamformer at the BS and the passive beamformer at the RIS, as shown on the right hand side of Fig. \ref{BB structure}.} For ease of illustration, let us consider a two-user setup, where the users are indexed by $m$ and $n$. The equivalent reconfigurable channels between the BS and each user result from the combination of the cascaded BS-RIS-user channels reconfigured by the RIS and the direct BS-user channel, as illustrated in Fig. \ref{BB structure}. \textcolor{black}{In other words, the equivalent reconfigurable channels depends on the configuration of the passive beamformer at the RIS.} Let user $n$ and user $m$ be the strong and weak NOMA users, respectively. User $m$ directly decodes its own signal received through the equivalent reconfigurable channel by treating the signal of user $n$ as interference. The achievable rate is denoted by ${R_{m \to m}}$. In contrast, user $n$ first decodes the signal of user $m$, at a decoding rate equal to ${R_{n \to m}}$, before decoding its own signal at an achievable rate equal to ${R_{n \to n}}$. In this context, it is worth mentioning that, for a given user decoding order, SIC can be successfully carried out at user $n$ provided that its decoding rate is no less than the achievable rate of user $m$. When compared to conventional networks, in RIS-NOMA networks, the optimal decoding order and the SIC performance depend on both the active beamformer at the BS and the passive beamformer at the RIS. Therefore, the joint optimization of the active and passive beamformers is very challenging.
%89

In this context, introducing the proposed equivalent reconfigurable channel illustrated in Fig. \ref{Beamformer-based} has the following advantages for the joint design of the active and passive beamformers. On the one hand, for a fixed passive beamformer at the RIS, we can map the RIS-NOMA network into a conventional network using the equivalent reconfigurable channel. Known efficient solutions for conventional NOMA networks can then be exploited for the optimization of the active beamformer for the proposed beamformer-based RIS-NOMA strategy. On the other hand, it has been shown that under some special channel conditions (e.g., quasi-degraded channels~\cite{quasi}), NOMA is capable of achieving the same performance as dirty paper coding (DPC). Based on this important insight, low complexity or even closed-form solutions for the optimization of the passive beamformer can be obtained by using the special channel conditions, where NOMA can achieve high performance, as the design objectives for the equivalent reconfigurable channel. \textcolor{black}{An initial study for applying the quasi-degradation condition as criterion for passive beamformer design can be found in~\cite{jianyue}. To further illustrate this concept, Fig. \ref{NDQ} compares the non-quasi-degradation (NQD) regions of two-user downlink multiple-input single-output (MISO) communication systems with and without RIS. For a given channel vector of user 1, the NQD region characterizes all possible channel vectors of user 2 which cannot satisfy the quasi-degradation condition, i.e., where NOMA performs worse than DPC. As can be observed, by reconfiguring the users' channels, the NQD region can be significantly reduced compared to a system without RIS. This confirms the effectiveness of the proposed beamformer-based RIS-NOMA strategy.} Nevertheless, the joint optimal NOMA decoding order and active and passive beamformer design is still challenging, especially for the dynamic RIS configuration, and requires further research.
%Additionally, in practice, RISs employ finite resolution phase shifters~\cite{Ding2020IRS}, and for passive beamformer design with discrete phases, efficient algorithms have to be developed and sophisticated mathematical tools have to be invoked.

\subsection{Cluster-based RIS-NOMA: Centralized and Distributed Design}
\begin{figure*}[!htb]
\centering
\subfigure[]{\label{centralized}
\includegraphics[width= 2in]{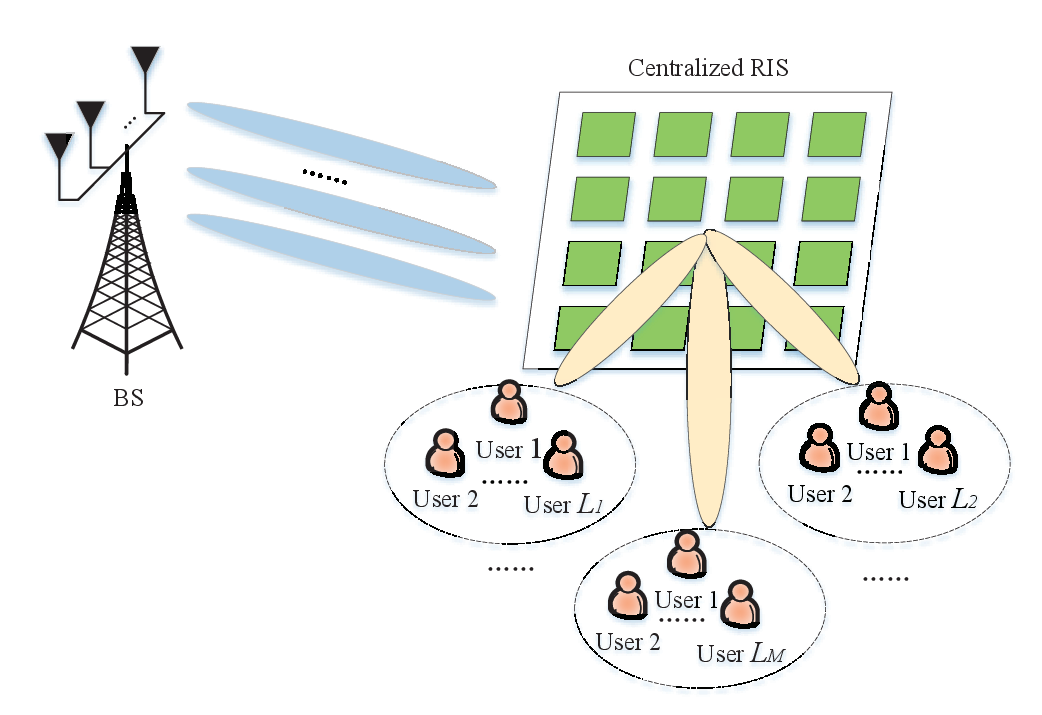}}
\subfigure[]{\label{distributed}
\includegraphics[width= 2in]{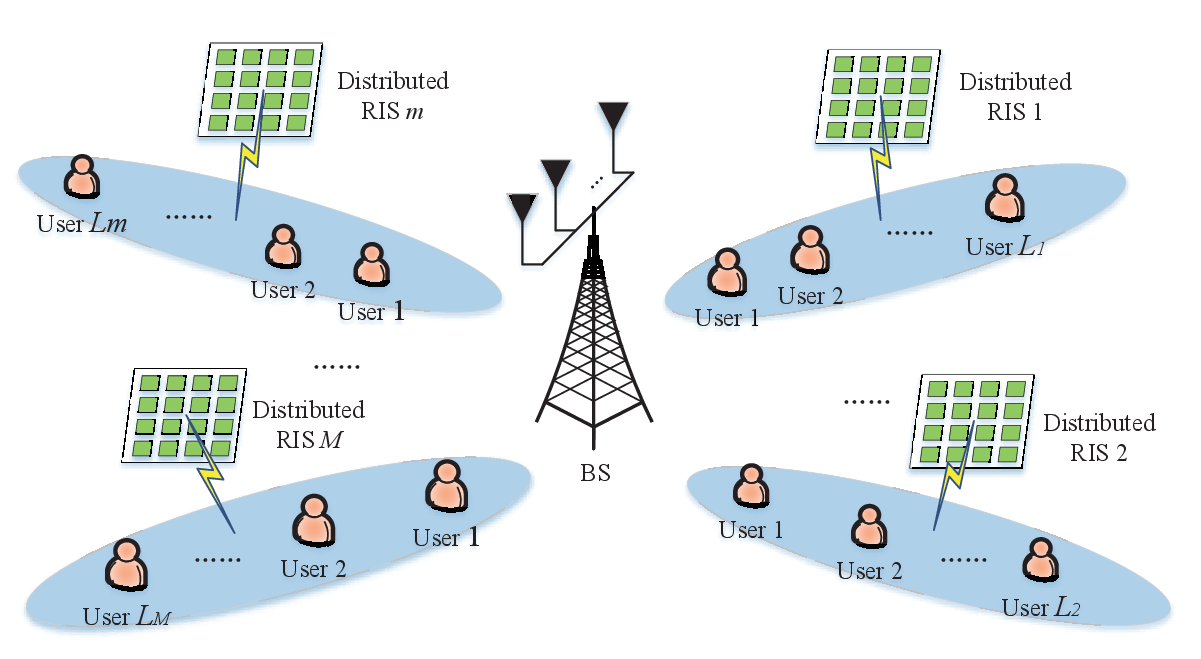}}
\subfigure[]{\label{result_center}
\includegraphics[width= 2in]{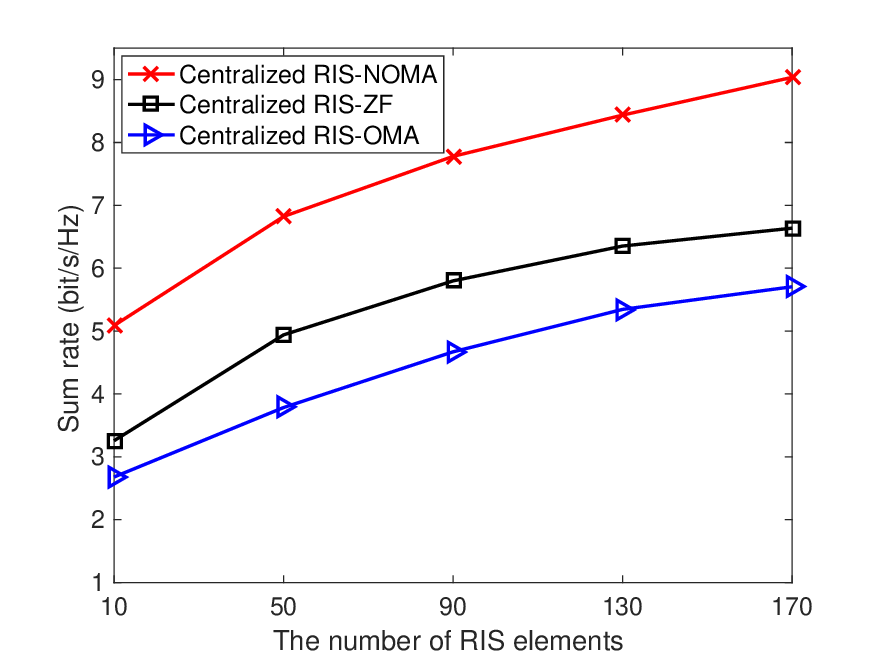}}
\caption{\textcolor{black}{Illustration of cluster-based strategies for RIS-aided multiple-antenna NOMA networks and initial numerical results. (a) A centralized RIS-enabled design. (b) A distributed RIS-enhanced design. (c) Performance gain of the centralized RIS-enabled cluster-based NOMA strategy. The system parameters can be found in~\cite[Section IV]{Zuo}.}}\label{Cluster}
\end{figure*}

The key idea behind the cluster-based RIS-NOMA strategy is to partition the users into several different clusters and serve the users in each cluster with one common beamformer. Conventionally, the MIMO NOMA channel can be decomposed into multiple single-input single-output (SISO) NOMA channels, by adopting transmit precoding at the BS and zero-forcing detection at each multiple-antenna user. However, such a decomposition relies on specific relationships between the number of equipped antennas at the BS and the number of users~\cite{Liu2018Multiple}, which limits the practicality of this cluster-based NOMA strategy. Fortunately, the deployment of RISs can relax the requirement for the aforementioned specific relationships since RISs are capable of providing additional passive array gains to facilitate operations, such as \emph{inter-cluster interference cancelation} and \emph{intra-cluster signal enhancement}. Inspired by this insight, we propose two designs for cluster-based RIS-NOMA, namely 1) A centralized RIS-enabled design; and 2) A distributed RIS-enhanced design, which are illustrated in Figs. \ref{centralized} and \ref{distributed}. In the following, we will compare the advantages and disadvantages of the two designs.
%49
%\begin{figure}[!htb]
%  \centering
%  \includegraphics[width=3.5in]{centralized.eps}\\
%  \caption{Cluster-based strategy for RIS-aided multiple-antenna NOMA networks: a centralized RIS enabled design.}\label{centralized}
%\end{figure}
\subsubsection{Centralized RIS-enabled Design}
A possible implementation of the centralized RIS-enabled design is illustrated Fig. \ref{centralized}, where the users are partitioned into several clusters \emph{enabled} by the passive beamformer at one centrally deployed RIS. The proposed centralized design for the cluster-based RIS-NOMA strategy is appealing for the case when there is no direct link between the BS and the users or when the BS-user links are highly correlated (e.g., in mmWave communications). In these scenarios, it is impossible to perform user clustering via the active beamformer at the BS, and the additional passive beamformer provided by the RIS comes to the rescue. \textcolor{black}{Fig. \ref{result_center} compares the performance of the centralized RIS-enabled cluster-based NOMA strategy with two baseline schemes for a mmWave communication system~\cite{Zuo}, where the direct BS-user link is assumed to be blocked. As can be observed, the proposed centralized RIS-enabled cluster-based NOMA scheme outperforms the baseline ZF-based and OMA-based schemes. As the number of RIS elements increases, the RIS gain for the NOMA scheme is more pronounced than the gains for the ZF and OMA schemes. This result underscores the advantages of the proposed centralized RIS-enabled cluster-based NOMA scheme.} Nevertheless, in the centralized design, an extremely large number of RIS elements may be required to realize a high passive array gain for user clustering. Furthermore, the passive beamformer design requires accurate instantaneous channel state information (CSI) of the entire network, which is non-trivial to obtain, especially when the number of users and the number of RIS elements are large. Therefore, the robust joint active and passive beamformer design for imperfect CSI is a promising research direction.
\subsubsection{Distributed RIS-enhanced Design}
In this design, multiple distributed RISs are deployed close to different user clusters which have been already separated by the active beamformer at the BS, as illustrated in Fig. \ref{distributed}. More particularly, each RIS aims to \emph{enhance} the performance of one specific user cluster within its own local coverage. The distributed RIS-enhanced cluster-based NOMA strategy is suitable when the users are widely distributed. In this case, the multiple distributed RISs can be deployed as `add-on' performance enablers for existing conventional NOMA networks. An advantage of this setup is that the impact of a given RIS on other unintended clusters is small due to the relatively large distances. Due to this property, the passive beamformer design at each distributed RIS depends mainly on the CSI of the users in its target cluster. However, the BS in this distributed design has to simultaneously exchange considerable amount of information with multiple RISs, which introduces new challenges especially when employing the dynamic RIS configuration. \textcolor{black}{An initial performance study for the distributed RIS-enhanced cluster-based NOMA strategy is reported in~\cite{Ding2020IRS}, where several distributed RISs were deployed to serve the cell-edge users, each of which was paired with one cell-center user to form one NOMA cluster. By doing so, both the maximum number of successfully served users and the coverage of the BS can be improved.}

\textcolor{black}{In Table \ref{table:structure}, we summarize the respective advantages and disadvantages of the proposed RIS-aided multiple-antenna NOMA transmission strategies, where ``BB-NOMA'' and ``CB-NOMA'' refer to beamformer-based NOMA and cluster-based NOMA, respectively.}
\begin{table*}[htb]
\caption{\textcolor{black}{Summary of RIS-aided Multiple-antenna NOMA Transmission Strategies.}}
\vspace{-1cm}
\begin{center}
\centering
\resizebox{\textwidth}{!}{
\begin{tabular}{|l|l|l|l|}
\hline
\centering
\textbf{Strategies}  & \textbf{Advantages} &\textbf{Disadvantages} & \textbf{Ref} \\
\hline
\centering
RIS aided BB-NOMA & \makecell[l]{High probability to achieve the same performance as DPC }  & Complicated joint optimal decoding order and beamformer design & \cite{jianyue} \\
\hline
\centering
Centralized RIS aided CB-NOMA&  \makecell[l]{High passive array gains provided by large-size RISs}  & Non-trivial CSI acquisition and limited coverage & \cite{Zuo} \\
\hline
\centering
Distributed RIS aided CB-NOMA&  \makecell[l]{Extended coverage and low-complexity beamformer design}  & \makecell[l]{Potentially high overhead for information exchange }    & \cite{Ding2020IRS}  \\
\hline
\end{tabular}
}
\end{center}
\label{table:structure}
\end{table*}%84

\section{Conclusions and Research Opportunities}
In this article, RIS-aided multi-user networks have been considered with a particular focus on the application of NOMA. Considering static and dynamic RIS configurations, the information-theoretic capacity limits and optimal RIS deployment strategies were discussed for RIS-aided single-antenna NOMA networks. It was shown that the dynamic RIS configuration is necessary to maximize the capacity gain and an asymmetric RIS deployment is preferable for NOMA as it facilitates significant channel differences between the users. For RIS-aided multiple-antenna NOMA networks, promising joint active and passive beamformer designs for different application scenarios were proposed for both beamformer-based strategies and cluster-based strategies. For beamformer-based strategies, a novel equivalent reconfigurable channel inspired design was advocated, which facilitates joint active and passive beamformer design by exploiting insights obtained for conventional NOMA networks. For cluster-based strategies, a centralized RIS-enabled design and a distributed RIS-enhanced design were developed, and their respective advantages and disadvantages were elaborated. Despite these recent advances, the design of future RIS-NOMA networks still holds many research challenges. Three particularly interesting problems are listed in the following:
\begin{itemize}
%  \item \textbf{Evaluation of Impact of Node Distribution Using Stochastic Geometry}: As a powerful mathematical tool, stochastic geometry is capable of capturing the spatial randomness of large-scale RIS-NOMA networks to provide practical RIS deployment guidelines. Since the reflection links depend on the spatial locations of the BSs, RISs, and users, performance evaluation is a difficult task. Additionally, multi-cell RIS-NOMA networks necessitate the redesign of the user association due to the flexible decoding order facilitated by RISs.
%  \item \textbf{Optimal Decoding Order and Beamformer Design}: For RIS-aided multiple-antenna NOMA networks, this problem becomes rather complicated since additional constraints imposed by the SIC decoding rate need to be satisfied, which causes the decoding order and the active and passive beamformers to be highly coupled. Therefore, determining the optimal decoding order and beamformer design for RIS-NOMA networks is a challenging task, which requires further research.
  \item \textcolor{black}{\textbf{Efficient CSI Estimation and Robust RIS-NOMA Design}: CSI acquisition is a particularly challenging issue due to the nearly passive nature of RISs. The accuracy of the CSI is more important for NOMA than for OMA since the CSI impacts both the user clustering scheme and the SIC decoding order. Therefore, efficient CSI estimation methods and robust RIS-NOMA design with imperfect CSI are two important future research topics.}
  \item \textbf{Machine Learning Algorithm Design}: In future ML-empowered RIS-NOMA networks, the design goals will be more ambitious than simply optimizing the network's capacity or delay, i.e., a single metric using single-component optimization. ML algorithm design is likely to probe further first by striking a trade-off between capacity, delay, and power, followed by finding all Pareto-optimal solutions for the resulting challenging multi-objective optimization problems.
  \item \textcolor{black}{\textbf{Hardware Realization and Experimental Verification}: This work focused on the theoretical aspects of RIS-NOMA networks. To validate the predicted performance gains, experiments have to be carried out. However, the hardware implementation presents several challenges, including the required real-time control and the inevitable hardware impairments affecting the RIS phase shift design and causing error propagation during SIC. These problems require further exploration in future research.}
\end{itemize}
%The reconfigurable channels facilitated by RISs enable a flexible decoding order design for RIS-NOMA networks. For RIS-aided single-antenna NOMA networks, the decoding order depends on the users' channel qualities, which are affected by the RIS passive beamformer.

\bibliographystyle{IEEEtran}
 \end{document}